# Software architecture for an unattended remotely controlled telescope

Robert J Lucas and Ulrich Kolb
Department of Physics and Astronomy
The Open University
Walton Hall
Milton Keynes
MK7 6AA

## *Abstract*

We report on the software architecture we developed for the Open University's remotely controlled telescope PIRATE. This facility is based in Mallorca and used in distance learning modules by undergraduate students and by postgraduate students for research projects.

## *Introduction*

PIRATE (Physics Innovations Robotic Astronomical Telescope Explorer) is a largely Open University funded facility consisting of a small aperture reflecting telescope on a robotic mount, in a robotic dome on top of the main observatory building at the Observatori Astronomic de Mallorca (OAM). Initially, the optical tube assembly (OTA) was a 14 inch (35 cm) f/10 Schmidt Cassegrain telescope (a Celestron 14; PIRATE Mark I). In August 2010 this was upgraded to a PlaneWave Instruments CDK17, a 17 inch (0.43 m) f/6.8 corrected Dall-Kirkham astrograph telescope (PIRATE Mark II).

PIRATE's main duties are for supporting research and distance teaching and it is the latter that brings quite stringent requirements for security whilst still offering the students comprehensive access and a rich experience. We cannot rely on the students closing the dome in bad weather, it must happen automatically. Additionally the observatory site is commonly unmanned during observing sessions, i.e. there is no one there to re-boot PCs, remove dust caps, attach dew shields, or make the telescope safe during bad weather. We have to achieve all these things without local assistance or remove the need for them in the first place.

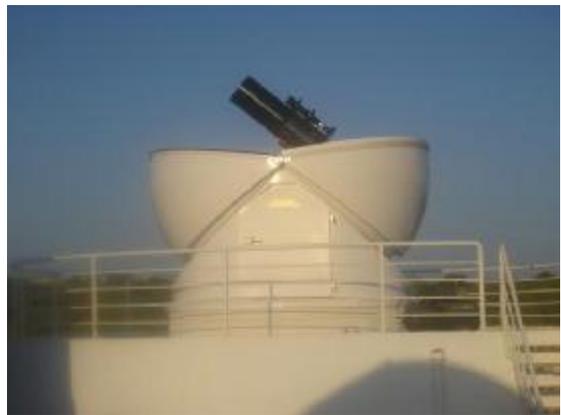

*A view of the PIRATE Mark I installation at the OAM*

## *The hardware components*

The hardware will be described pointing out the features where relevant that impact on the remote control aspects of PIRATE. The hardware consists of the following devices:

- A 3.5 m AllSky dome made by Baader Planetarium[1]. This is of the clamshell design, so there is no dome slit that needs to be rotated in order to keep a celestial target in view. The dome is normally powered from the mains but it has a battery back-up which is essential for closing the dome during a power cut. It is equipped with its own temperature sensors for ambient and cloud as well as a rain sensor.

- A PC running Windows XP known as PC1. This hosts all the necessary software for controlling all aspects of the telescope. Some of this software is bespoke and its design will be discussed in this paper.

There is also a great deal of software that is "off the peg", such as ACP[2] for executing imaging plans and MaxIm DL[3] for camera control and imaging.

- A backup PC also running Windows XP and known as PC2. This is used for monitoring weather from the non-dome sources of weather information and for a backup for PC1 should that computer develop a fault.

- A German Equatorial Mount (GEM), this is a Paramount ME[4], an extremely robust mount that has been designed with remote operation in mind. This is connected to the PC via a serial port.

- A Celestron[5] optical tube assembly (PIRATE Mark I) as described above. This is attached to the mount and has no electrical connections.

- A guide 'scope consisting of an inexpensive Celestron refractor. This also has no electrical connections.

- A main camera, SBIG STL-1001E[6], with a controllable Peltier cooler, for long exposure, high resolution imaging and photometry. This is coupled to a filter-wheel containing eight filters. The camera and filter-wheel connect to the PC via USB.

- A guiding camera, SBIG ST-402ME[6], with controllable Peltier cooler, for accurate guiding of the mount. This takes a very small image of a suitable star every few seconds, the centoid of the star is calculated in MaxIm DL and depending on how the star wanders, various guiding alterations are sent to the mount to compensate.

- An Optec electronic focuser[7] (for the PIRATE Mark I OTA) with temperature compensation. This is used with FocusMax[8] to achieve as near perfect focus as possible through all the different filters. This is connected to the PC via a serial port.

- Heating strips for the main OTA and the guide 'scope. These are to prevent dew build up on the corrector plate of the PIRATE Mark I OTA and the objective lens of the guide-scope.

- Two banks of four Phidgets[9] which are PC controllable mains power switches which we use for powering up much of the hardware described above.

- Several webcams giving views of PIRATE from inside and outside the dome. These are IP enabled and can be accessed directly from the internet as well as from the PIRATE web pages[10].

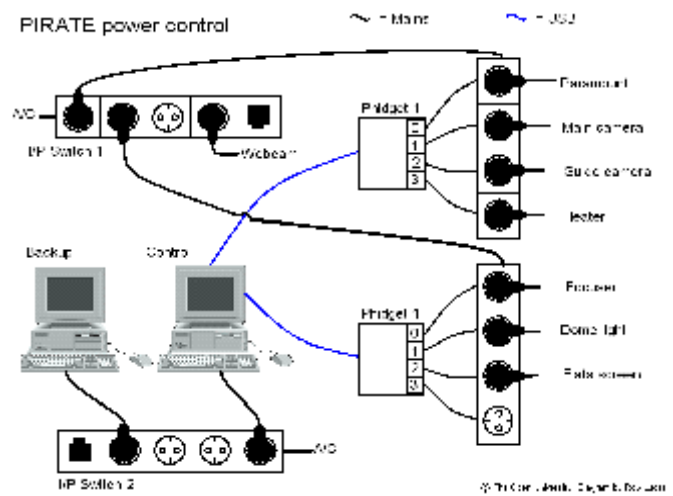

*Representative view of connections showing how power is supplied to the various PIRATE hardware components.*

## *The off-the-peg software*

**The Sky**

This is a planetarium style program by Software Bisque[4] (the manufacturers of the Paramount) that also doubles as the driver for the Paramount GEM. It gives a view of the present state of the heavens and where the telescope is currently pointing.

**MaxIm DL**

The imaging software, MaxIm DL[3], controls both cameras and the filter wheel. As well as controlling exposure times and guiding it also provides the means to cool the cameras to particular temperatures. Not only does it do these functions via a visible interface but it also exposes a number of ASCOM[11] programmable objects, such as camera and document objects that can be used by other programs such as the main plan controller, ACP. But as importantly, we can use these in our own programs.

### T-Point[12]

This application works in conjunction with The Sky[4], which it is supplied with, to improve polar alignment and improve pointing by mathematically modeling all the various factors that contribute to pointing errors such as tube flexure. It does this by using a series of observations where operator chosen stars are driven to by The Sky driving the mount and then the operator manually applies a correction to the position to centre the star in the field of view whether through an eyepiece or on a CCD chip. Clearly, one of the factors that can introduce a pointing inaccuracy is any error in polar alignment, hence T-Point can advise on what corrections to make in Alt -Az to improve the mount alignment. In the case of the Paramount it makes very precise recommendations in terms of the number of turns to make to the Alt-Az knobs that control the precise alignment of the mount.

### Pin Point[2]

This uses a database of 2 million stars to match against any image. This is used to correct for any pointing errors by identifying the matched stars in the field and then calculating the necessary movement to be applied to the mount to correct for any error. In practice this means that we can guarantee that our targets are perfectly centred on our imaging chip.

### FocusMax[8]

FocusMax is a freeware application that allows excellent automated focusing. It calibrates the focus control by taking a series of exposures at different focus positions and plotting these positions against the half-flux diameter of a stellar image. It then calculates the lines of best fit, which it calls the V-curve. The obtained V-curve can then be used to compute the ideal focus position given the half-flux diameter of a star image. Different V-curves are calculated for all the filters.

### ACP[2]

This is the main controlling application that coordinates the use of the above software components. Principally it allows the execution of observing plans and the remote operation via a web interface. However, in achieving these main aims it provides a myriad of functions from automated flat taking to dome control and integration of weather data.

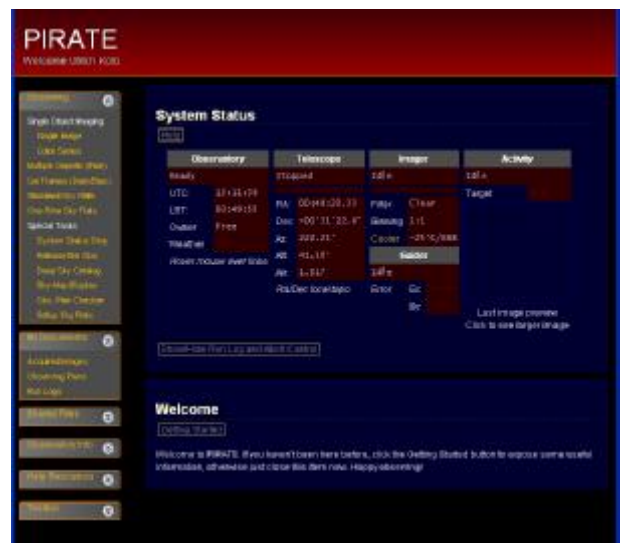

*The ACP web user interface*

### VNC[13]

This provides us with remote access to the control PC at the observatory via the internet. When executed from any other PC, which can be anywhere in the world, it presents a scrollable window that is the screen of the control PC. The computer can then be controlled by the keyboard and mouse of the remote machine exactly as if you were sitting in front of it.

## *The bespoke software*

**Powering the hardware devices, the SwitchServer**
From early on, it was clear that we needed to be able to switch the electrical supply to all devices from the control PC; we would then have remote access to powering these devices via VNC. One of the authors had some experience of using Phidgets to control switching of hardware for stereoscopic use and a check of their specifications verified that they would be within their capabilities for switching mains circuits. Initially a simple Phidget Switcher was coded that gave us the ability to switch on and off devices by using this program on PC1, which we could then use remotely via VNC. However, we knew that this was not going to be a viable solution for any students using PIRATE as we did not want them to have any direct access to PC1 as this might cause any number of security issues. Therefore we needed to investigate how remote access to the switches could be provided.

The Phidgets come with software for remotely controlling them over the internet and even had facilities for setting and querying variables that might have been used for controlling other aspects of PIRATE.  Unfortunately these remote facilities could not be used as they required multiple TCP/IP connections using addresses that were not compatible with the Open University's firewall restrictions.  Hence it was necessary to build our own SwitchServer that used a single TCP/IP connection with an address and port number of our choosing.

It was clear that the dome would need to be controlled remotely.  We were involved from the earliest stages with the design and implementation of the AllSky ASCOM compliant dome driver. The dome has its own temperature and rain sensors.  We had made the design decision that the firmware that controls the opening and closing of the dome leaves via a set of four motors and position sensors should be as autonomous as possible.  So it was decided that should the dome rain sensor detect rain then the firmware should close the dome without any instruction being required from the driver.  It was also decided that should the firmware not be polled by the driver for more than a few minutes then it should assume that the connection to the PC had failed and in this case the firmware should close the dome.  We believe that this makes the entire system considerably safer as we do not need to rely on the controlling PC issuing the necessary command in response to the conditions.  The chain of command between the detection of the rain and the closure of the dome is as short as it can be.
The SwitchServer can issue open and close directives by use of the ASCOM dome object exposed by the driver.
Additionally the SwitchServer obtains the current ambient temperature from the dome-mounted sensor; this is subsequently used to determine target temperatures to cool the cameras to.  The cameras need to be cooled to as low a temperature as is feasible to minimize thermal noise in the electronics.  This needs to be done in stages to prevent condensation building on the chips.  First the cameras are connected and the coolers switched on.  Then an initial setpoint is chosen as 5 degrees below ambient.  When this is reached there is a programmed pause before issuing the next setpoint.  This continues until the final setpoint (as calculated from the ambient temperature) is reached.  Finally the power that is being provided to the Peltier cooler is monitored, if this is close to 100% then the final setpoint is increased by 5 degrees to ensure that the temperature can be maintained.  Although different target temperatures are used for each camera, the algorithm described above is essentially the same.

The SwitchServer also has access to sunrise and sunset times and will close the dome should it find it open during daylight and not allow it to be opened in daylight.  This is because the OTA has no cap so that if the dome were to be open during the day then it is possible that the Sun might shine down the tube causing the cameras to become very hot and probably sustain damage.

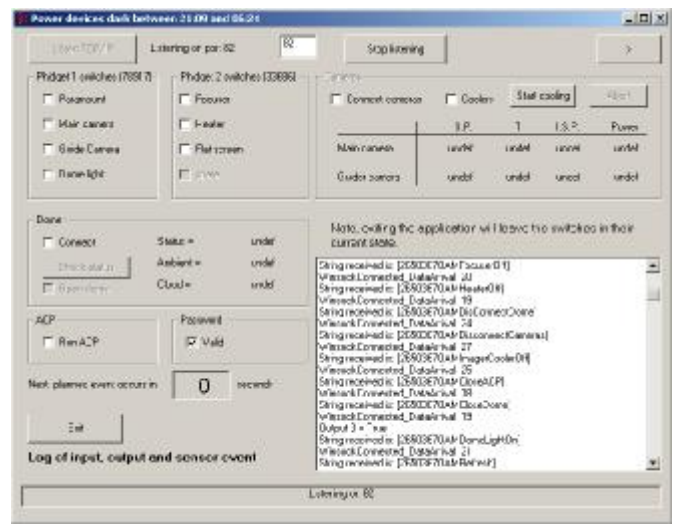

*The SwitchServer running on the control PC*

The SwitchServer can be used directly by logging onto PC1 locally or remotely.  For remote student access it uses a single TCP/IP socket to listen for incoming connections.  On receiving a connection request it establishes the connection on a second socket and waits for the command string to arrive via the data received event associated with the socket.  The request is parsed, the associated password is validated, and finally the command is passed to the relevant object (which is usually either the dome, phidget or camera object).  The above screen shot of the SwitchServer shows the command strings being received in the large scrolling window at the bottom right of the application.

**The SwitchClient**
The SwitchClient sends commands to the SwitchServer.  The user needs to provide a password and this is concatenated with each command sent so that it can be validated by the SwitchServer.  A program has been written that generated passwords that are valid from midday of the night in question until midday on the following day. Typically a different group of students use PIRATE on each night until all of the groups have had their turn and then they start over. Each group is issued with a password that is valid for their particular session preventing one group interfering with the observations of any other.

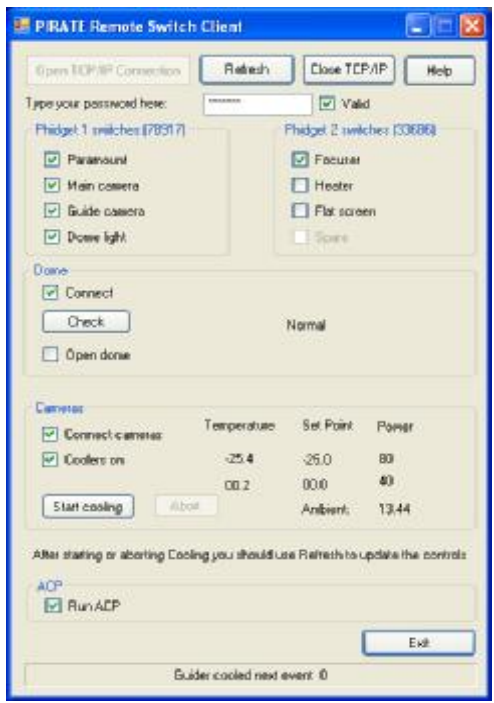

*The SwitchClient showing the cooled cameras and ACP having been started.*

The SwitchClient connects to the TCP/IP socket on the SwitchServer for each command issued. This means that several different users can use the SwitchClient at the same time. Typically the students in an observer team will nominate one to actually submit the commands whilst the others will be able to watch the current state by simply clicking on the Refresh button. When the command has been executed on PC1 a message string is returned via TCP/IP to the SwitchClient that contains the status of all of the checkboxes and writeable labels that you can see in the picture of the SwitchClient given above.

## *Evaluation*

PIRATE Mark I started its commissioning phase in summer 2009 and went live with its first set of Open University (OU) students on 17th April 2010. For the most part it has performed well with very little down time, and the student response was very positive.

In order to maximize the chance for the cohort of 25 students to experience good observing conditions during the run-time of their OU module, a tight observing schedule of 40 consecutive nights was drawn up. With hindsight, we should have given ourselves one evening per week to perform upgrades and testing.

The intensive PIRATE use by OU students represented also a demanding road-test for the ASCOM-based AllSky dome driver we developed for Baader Planetarium. This resulted in several improvements in the dome firmware as well as the design of the dome driver. One example is the separation of higher-level tasks such as polling and status queries which are now executed by the SwitchServer, from sending basic operational commands

Dealing with these teething issues meant that on occasion the (UK-based) night duty astronomer had to step in and take a higher level of control on PC1 via VNC. In over 18 months of operations, the only two occasions when local on-site OAM assistance was needed during the night was when the OAM internet service provider conducted unannounced upgrades, and when the un-interrupted power supply (UPS) unit initially used in the dome failed. The UPS is now known to have been underpowered, and a more tailored solution is being developed by Baader Planetarium.

## *Improvements*

Further improvements are continuously being implemented, for the benefit of future student cohorts. The SwitchServer now monitors the internet for communication problems, and it has access to the humidity and wind speed sensors that are not part of the dome's setup but are currently monitored on the second PC. We are considering replacing the SwitchClient with a web interface, served by the SwitchServer. The dome driver can control the four dome leaves independently, so as to provide a shelter for the telescope in case of higher wind speeds. This functionality is not yet available from within the SwitchServer, and hence unavailable to student users.

## *Conclusions*

When we started this project we thought that very nearly all of the requisite software and hardware would be available off the shelf. This has not been our experience. We have found ourselves climbing a steep learning curve where, because of the security issues involved in giving students access to an unmanned telescope, we have had to provide a great deal of the controlling software which has needed a good deal of thought to enable us to negotiate various restrictions whilst providing the needed functionality.  Ultimately we have achieved what we set out to do.  We have had a very successful first cohort of students using PIRATE and most problems that have arisen have been swiftly resolved.  We hope that our experiences will be of value to others who wish to provide similar facilities.

## *Acknowledgements*

We gratefully acknowledge the contribution of OU and OAM staff and affiliates who have contributed to the success of PIRATE. Vadim Burwitz and Stefan Holmes played major roles.